\documentclass{article}

\usepackage[english]{babel}
\usepackage{authblk}
\usepackage{cite}

\usepackage[letterpaper,top=2cm,bottom=2cm,left=3cm,right=3cm,marginparwidth=1.75cm]{geometry}

\usepackage{siunitx}
\usepackage{amsmath}
\usepackage{graphicx}
\usepackage{amssymb}
\usepackage[colorlinks=true, allcolors=blue]{hyperref}

\title{An array of bulk-acoustic-wave sensors \\ as a high-frequency antenna for gravitational waves}

\author[1,2]{G.~Albani}
\author[1,2]{M.~Borghesi}
\author[1,2]{L.~Canonica} 
\author[1,2]{R.~Carobene}
\author[1,2]{F.~De~Guio}
\author[1,2]{M.~Faverzani}
\author[2]{E.~Ferri}
\author[1,2]{R.~Gerosa}
\author[1,2]{A.~Ghezzi}
\author[1,2]{A.~Giachero}
\author[2]{C.~Gotti}
\author[1,2]{D.~Labranca}
\author[1]{L.~Mariani}
\author[1,2]{A.~Nucciotti}
\author[2]{G.~Pessina}
\author[1,2]{D.~Rozza}
\author[1,2]{T.~Tabarelli de Fatis}
\affil[1]{Universit\`a degli Studi di Milano Bicocca, Milan, Italy}
\affil[2]{INFN, Sezione di Milano-Bicocca, Milan, Italy}

\newcommand{\acapo}{\vspace{0.2cm} \par}
\begin{document}
\maketitle

\begin{abstract}
In their simplest form, bulk acoustic wave (BAW) devices consist of a piezoelectric crystal between two electrodes that transduce the material's vibrations into electrical signals. They are adopted in frequency control and metrology, with well-established standards at frequencies of 5~MHz and above. Their use as a resonant-mass strain antenna for high-frequency gravitational waves has been recently proposed (Goryachev and Tobar, 2014). The estimated power spectral density sensitivity at the resonant frequencies is of the order of $10^{-21}\, \textrm{strain}/\sqrt{\textrm{Hz}}$. 
In this paper, after introducing the science opportunity and potential of gravitational wave detection with BAWs, we describe the two-stage BAUSCIA project plan to build a multimode antenna based on commercial BAWs, followed by an optimized array of custom BAWs. We show that commercially available BAWs already provide sensitivity comparable to current experiments around 10~MHz. Finally, we outline options for optimization of custom devices to improve sensitivity in an unexplored region, probe multiple frequencies between 0.1 and 10 MHz, and target specific signals, such as post-merger emission from neutron stars or emission from various dark matter candidates.  
\end{abstract}

\section{Introduction}

The direct observation of gravitational waves (GW) by interferometric experiments~\cite{PhysRevLett.116.061102} marked the beginning of a new era of astrophysical investigation, which makes it possible to test gravitation and set new constraints on astrophysical models. The search for correlations between gravitational and electromagnetic signals has promptly become a new standard that enabled, for example, the first measurement of the expansion parameter of the universe using gravity alone~\cite{Abbott2017}. Current and future interferometers, on the ground or in space, will mark a significant step forward in studying gravitational waves. The expected sensitivity of planned large-scale projects~\cite{Moore_2015} will consolidate the knowledge of massive and supermassive GW sources and expand the observable horizon into the high-redshift deep past. However, these projects only target gravitational signals at frequencies lower than approximately 1~kHz, with a cut-off essentially set by the length scale of the apparatus. 

\acapo Theoretical motivations for potential sources of high-frequency gravitational waves (HFGW) have been steadily increasing recently~\cite{Aggarwal_2021}. In the region between 0.1 and 10~MHz, corresponding to the sensitivity of the experimental technique discussed in this paper, notable examples include GW emission from dark matter candidates, like mergers of planetary-mass primordial-black-hole binaries~\cite{PhysRevD.106.103520} or axions collapsing into a massive black hole~\cite{PhysRevD.83.044026}, and from conventional processes, like QCD phase transitions following the event merger of neutron-star binaries~\cite{casalderreysolana2022}. 
Therefore, an apparatus with sensitivity in this frequency range is well-motivated. It would be both complementary and supplementary to large-scale interferometers: GW detection in this range would indicate the existence of non-conventional sources or confirm the observation of neutron-star mergers in coincidence with GW interferometers.  
\acapo
In parallel with the growth of theoretical motivations, many experimental methods have been explored and shown to exhibit sensitivity to HFGWs at different frequencies, from 100~kHz to above 1~GHz. These include small-scale interferometers~\cite{Cruise_2006, Cruise_2012, PhysRevD.77.022002, PhysRevLett.101.101101, PhysRevD.95.063002}, dissimilar techniques exploiting the coupling between gravitational and electromagnetic fields\cite{Ejlli2019, PhysRevD.102.103501, Ito_2020, Herman_2021, Tobar_2022, Domcke_2022, PhysRevD.110.023018}, gravitational shifts of the Mössbauer absorption \cite{GAO20242795},  and various resonant-mass detectors~\cite{domcke2024magnetsweberbargravitational,PhysRevLett.110.071105, PhysRevD.90.102005, tobar2024G}. Bulk-acoustic wave (BAW) devices belong to the latter category. They consist of a piezoelectric crystal between two electrodes that converts the acoustic waves within the material into electrical signals. They are adopted in frequency control and metrology, with well-established standards at frequencies of 5 MHz and above, and give one of the best levels of frequency stability, with $Q$-factors of the order of 10$^6$ at room temperature. 
\acapo
Following a seminal paper by Goryachev and Tobar~\cite{PhysRevD.90.102005}, the use of BAWs as a resonant-mass strain antenna has been developed and successfully brought into operation at the University of Western Australia (UWA)~\cite{PhysRevLett.127.071102, SciRep13-10638}. For GW detection, BAWs act as acoustic resonant cavities with high sensitivity to impinging strain fields with frequencies matched to the cavity vibration modes. The electric signals generated by crystal vibrations are read out by a superconducting quantum interference device (SQUID) amplifier coupled to the quartz resonator electrodes. The system operates at cryogenic temperatures ($T< 4$~K), where the thermal noise is reduced and the Q-factor of the cavities increases. The latest setup, dubbed Multimode Acoustic Gravitational Wave Experiment (MAGE)~\cite{SciRep13-10638}, features two identical BVA-type quartz resonators coupled to two independent SQUID amplifiers held at a constant temperature of 4~K. Each detector is sensitive to gravitational radiation in multiple narrow bands corresponding to the crystal’s overtone modes between 5 and 10~MHz, which are simultaneously monitored. The projected peak spectral strain sensitivity, expressed in terms of single-sided power spectral density, for operation at 20~mK is 5--10$\times 10^{-22}\, \textrm{strain}/\sqrt{\textrm{Hz}}$. 
\begin{figure}[thb]
    \centering
    \includegraphics[width=0.75\linewidth]{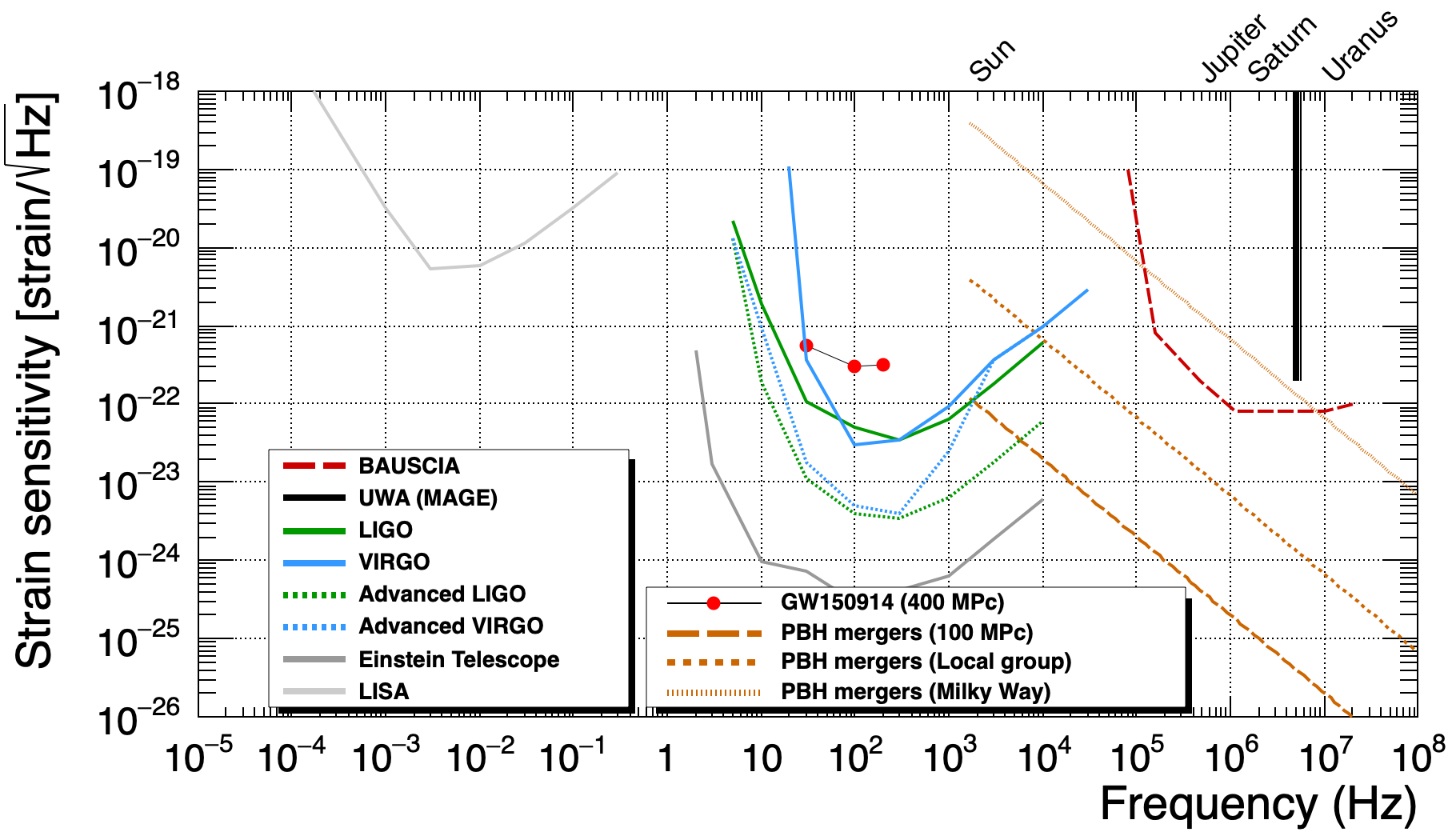}
    \caption{Sensitivity of a current BAW antenna (black line)~\cite{SciRep13-10638} and potential coverage of the proposed array of BAWs (red line) compared to the sensitivity of existing and planned interferometers (adapted from Ref.~\cite{Moore_2015}) and putative GW radiation from planetary-mass primary-black-hole mergers.  
    }
    \label{fig:sensitivity}
\end{figure}

\acapo This paper describes the BAUSCIA project (Milan's transliteration of BAWSHA for Bulk Acoustic Wave Sensors for High-frequency Antennas). In a two-stage plan, we aim to build a multimode antenna based on commercial BAWs for multi-site detection in coincidence with MAGE, followed by an optimized array of BAWs to sample multiple frequencies between 100 kHz and 10 MHz. After a description of the experimental approach (Section~\ref{Sec1}), we discuss the characterization of some commercially available devices, which turn out to be suitable for an antenna with sensitivity and frequency coverage comparable to MAGE (Section~\ref{Sec2}). Dedicated effort into research and development will improve the sensitivity and extend coverage to frequencies beyond those of direct interest to commercial applications (Section~\ref{Sec3}). For illustration, Fig.~\ref{fig:sensitivity} compares the signal strength of putative GW sources with the envelope of the projected single-sided peak strain sensitivities of an array of tens of BAWs, estimated in Section~\ref{Sec3}. The figure also shows the complementarity of high-frequency GW detectors to the existing or planned large-scale interferometers, which cover a disjoint frequency interval.

\acapo
In addition to high-frequency gravitational waves, this apparatus can probe various dark-matter scenarios, including ultralight scalars~\cite{PhysRevLett.116.031102, PhysRevLett.124.151301} or dark photons \cite{trickle2025piezoelectricbulkacousticresonators}. The sensitivity depends approximately on the same BAW parameters as for GW searches. Bulky piezoelectric crystals with resonant frequencies in the 10-100 kHz range could evade current constraints. 

\section{Detection concept and setup}
\label{Sec1}

\begin{figure}[hbt]
    \centering
    \raisebox{0.93cm}{\includegraphics[width=0.55\linewidth]{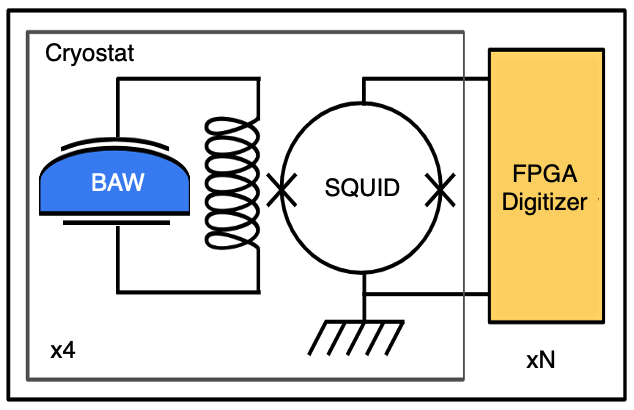}}
    \includegraphics[width=0.44\textwidth]{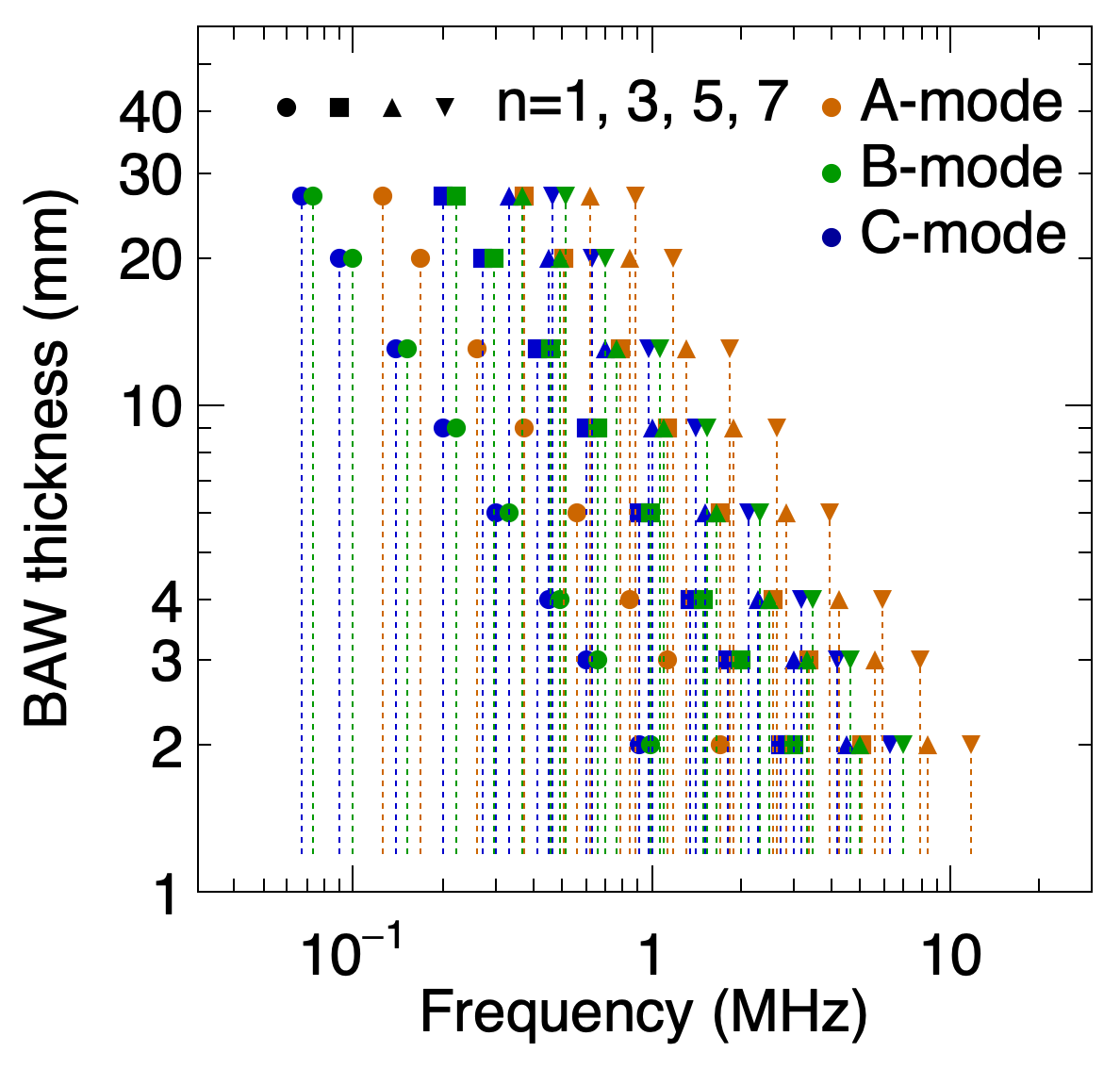}
    \caption{Sketch of the BAUSCIA module prototype (left panel) and map of the resonant frequencies of an array of eight quartz BAWs of different thicknesses (right panel)}. 
    \label{fig:setup}
\end{figure}

The instrumental technique adopted by the BAUSCIA project is based on the pioneering work that led to the MAGE experiment. A key objective of BAUSCIA is the construction of an apparatus with sensitivity across more than two decades of frequencies to offset the narrow-band response limitation of the resonant-cavity approach to GW detection. To this end, the apparatus is designed for scalability. The left panel of Fig.~\ref{fig:setup} sketches the baseline modular structure and the readout chain of the prototype under development. Each module features four BAW sensors coupled to independent SQUID amplifiers, matching the modularity of the prototype back-end system. An external veto against cosmic rays will be considered to exclude potential sources of spurious events, as those observed in~\cite{PhysRevLett.127.071102}.

\acapo An impinging gravitational wave causes a driving force on the BAW mass elements proportional to the Riemann tensor perturbation from the wave: $(d^2x^j/dt^2)=(R^j_{0k0})x^k$. For a GW-driven resonator, the equation of motion for the displacement $u_\lambda(\textbf{x},t)=a_\lambda(t)U_\lambda(\textbf{x})$ of a stationary mode $\lambda$ reads: 
\begin{equation}
\ddot{a}_\lambda+\gamma_\lambda\dot{a}_\lambda+\omega^2_\lambda a_\lambda = -c^2 R_{0i0j}\int_V \frac{\rho}{m_\lambda} U^i_\lambda(\textbf{x},t)x^j d\nu
    \label{Eq:GW-drive}
\end{equation}
where $\gamma_\lambda$ and $\omega_\lambda=2\pi f_\lambda$ are the mode width and angular frequency, $\rho$ is the mass density of the BAW, and $m_\lambda = \int_V U_\lambda^2 \rho d\nu$ its effective mass or, in other words, the mass that participates in the vibration. The integral over the BAW volume on the right-hand side of Eq.(\ref{Eq:GW-drive}) quantifies the coupling between the impinging GW and the acoustic mode of the cavity~\cite{Goryachev_2014}. Quartz cavities support vibrations with longitudinal and transverse polarizations and different velocities $v_A = 6757$~m/s (longitudinal), $v_B = 3966$~m/s (fast shear), and $v_C = 3611$~m/s (slow shear). The crystal thickness and the phase velocity set the frequency spectrum of the modes. Only odd overtones exhibit a piezoelectric counterpart that gives rise to a voltage difference between the top and bottom electrodes, yielding a rich spectrum of audible resonant frequencies (see next Section for further details):
\begin{equation}
\label{eq:modes}
    f_{n,X} = n \frac{v_X}{2d}, \;\;\;\;\ (n=1,3,5,... ; X=A,B,C)
\end{equation}
This combination of modes and overtones allows a single BAW to sample multiple frequencies over a broad range. All the BAWs of the current prototype have a thickness $d=1$~mm and are optimized for clock applications at $f \simeq 5$~MHz, corresponding to the third overtone of the slow-shear $C$-mode. An array of BAWs of different thicknesses will enable sampling a broad spectrum of frequencies, as illustrated in the right panel of Fig.~\ref{fig:setup}. 

\acapo The signal from the BAW resonators is read out using Magnicon DC-SQUIDs with a nominal input impedance of 400~nH and 20/50~MHz FLL/open-loop bandwidth. The BAW sensors and the readout SQUID amplifier are operated at cryogenic temperature, in a dilution refrigerator that can reach a base temperature of 20~mK. The back-end data acquisition, outside the dilution refrigerator, adopts field-programmable-gate-arrays (FPGA) waveform digitizers\footnote{Xilinx Zynq UltraScale+ FPGA RFSoC4x2, which supports four 14-bit 4.096 Gsample/s ADCs} to continuously and simultaneously monitor all the relevant resonant modes on a single readout line through digital lock-in amplification. This system provides the flexibility to adapt to hardware changes, chiefly the BAW thickness and corresponding resonant frequencies.
For scalability, we are also considering an alternative layout with multiple (two or four) BAWs in parallel at the input of a single SQUID, to overcome the inadequate availability of SQUIDs on the market. Based on previous work~\cite{PhysRevD.90.102005, SciRep13-10638}, we expect the sensitivity of each independent readout line to be limited by the BAW thermal noise at resonance. Multiple BAWs with narrow resonances at different frequencies on the same readout line would remain electrically and mechanically decoupled (except for a change in the input impedance to the SQUID) and only add off-resonance, damped noise. The results of this layout will be the subject of a future report. 

\acapo The peak sensitivity at the resonant frequencies is described by the single-sided power spectral density of the Nyquist thermal noise produced by the quartz resonator near an acoustic mode~\cite{PhysRevD.90.102005, SciRep13-10638}: 
\begin{equation}
\label{eq:sensitivity}
    S^{+}_{h}(\omega_{\lambda})=\frac{2}{\bar{\xi}_{\lambda} d}
\sqrt{\frac{k_bT_{\lambda}}{m_{\lambda}Q_{\lambda}~{\omega_\lambda}^3}} \ \ \ [\textrm{strain}/\sqrt{\textrm{Hz}}],
\end{equation}
where $k_b$ is the Boltzmann constant, 
$T_\lambda$, $m_\lambda$, and $Q_\lambda$ are the mode temperature, effective mass, and quality factor. 
The term $\bar{\xi}_{\lambda}d/2$, where $d$ is the BAW thickness, is a shorthand notation for the integral on the right-hand side of Eq~(\ref{Eq:GW-drive}), which expresses the coupling between the impinging GW and the acoustic mode of the cavity. The unit-less coupling parameter $\bar{\xi}_{\lambda}$ is inversely proportional to the square of the overtone number ($\bar{\xi}_{\lambda} \propto 1/n^2$) through a coefficient of the order of 1, which accounts for the trapping of the vibrational energy within the BAW volume~\cite{PhysRevD.90.102005}. For well-trapped modes, the effective mass scales as $m_\lambda \propto 1/n$ (see next section). As a consequence, net of quality factor variations with the overtone, the noise power-spectral-density increases with the overtone as $S^{+}_{h}(\omega_{\lambda}) \propto n$. In other words, a given cavity provides optimal sensitivity at the lowest overtones of its resonant modes ($n\lesssim7$). Figures of merit of a BAW-based antenna at a resonant frequency are a low operating temperature (cryogenic), a high effective mass, and a large quality factor. All the relevant parameters depend on the BAW geometry and the operating temperature and must be characterized experimentally. 

\section{Bulk acoustic wave cavities}
\label{Sec2}

\begin{figure}[thb]
    \centering
    \includegraphics[width=0.95\linewidth]{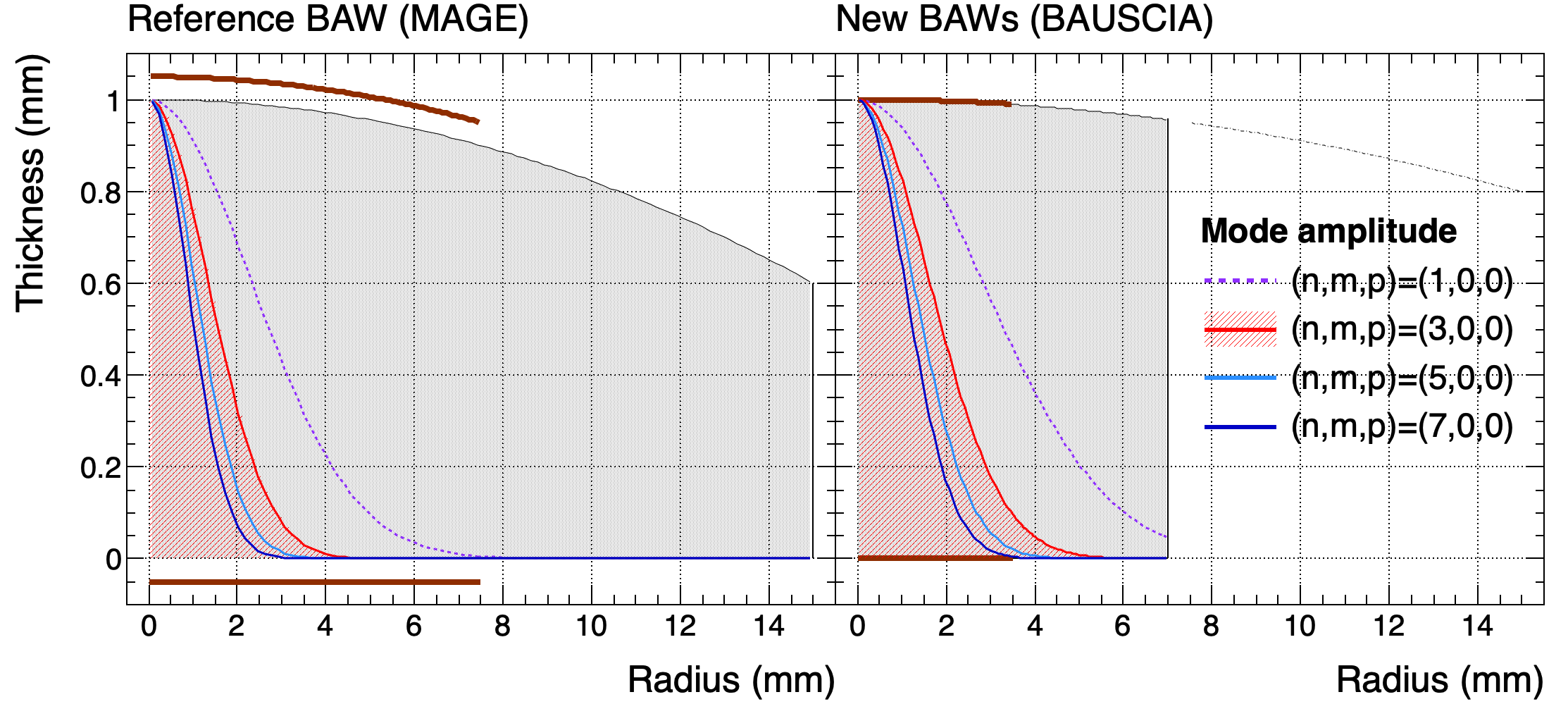}
    \caption{Cross section in the $(z,r)$ plane of half the radial extension of the reference (left) and new bulk acoustic wave samples (right). The lines show the amplitude profile of the modes in the transverse plane for $\lambda = \{X,n,0,0\}$ (see text for details).}
    \label{fig:Modes}
\end{figure}

A bulk acoustic wave cavity consists of a plate of piezoelectric material specially designed to support acoustic vibrations with high quality factors. Ideally, the vibrating part of the plate is mechanically isolated from the environment to minimize any energy leakage. A typical realization is a thin disk of radius $L$ and thickness $d \ll L$ with plano-convex surfaces, anchored by rigid clamps from the sides. The surface curvature traps the vibrations in the centre of the cavity, reducing the dissipation through the clamps~\cite{Tiersten86, Galliou2013}. 
For effective excitation and pick-up of the vibrations, the plate is sandwiched between two electrodes, either deposited on the surface or separated from it by a thin vacuum gap. The latter configuration, known as BVA~\cite{1537081}, further isolates the vibrating mass from the environment.
Figure~\ref{fig:Modes} shows a schematic cross section of half the radial extension and the transverse distribution of trapped phonons for some acoustic modes, for the two sets of BAW samples characterized in this work. Their geometries are compared in Table~\ref{tab:my_label}. 

\begin{table}[htb]
    \centering
    \begin{tabular}{|c|c|c|c|c|c|} \hline
 Sample & $d$ (mm) & $L$ (mm) & $R_c$ (mm) & $\sigma_1$ (mm) & electrode \\ \hline \hline
 Reference & 1 & 15 & 300 & 2.3 & separated \\
 New       & 1 &  7 & 600 & 2.8 & deposited \\ \hline
    \end{tabular}
    \caption{
Properties of the reference and new BAW cavities: thickness, $d$; radius, $L$; radius of curvature of the convex surface, $R_c$; and effective radius of the phonon distribution, $r_1$, for the overtone number $n=1$ (see text for details).}
    \label{tab:my_label}
\end{table}

\acapo The reference sample (courtesy of M. Tobar and collaborators) is identical to the BAWs used in the MAGE experiment. It consists of a 1 mm-thick, 30 mm-diameter, electrode-separated disk, manufactured by Oscilloquartz SA~\cite{Galliou2013}. The new samples are 1 mm-thick, 14 mm-diameter disks, from the Rakon Ltd company stock, with 7 mm-diameter and 200 nm-thick gold electrodes deposited on the disk surfaces (Fig.~\ref{fig:Rakon}). All the resonators, enclosed in a vacuum-tight case, are doubly-rotated stress-compensated (SC) cut plates, realized from high-purity crystal quartz. They are optimized for frequency control operations at $f=5.175$~MHz, corresponding to the third overtone of the slow shear mode ($C3$), and provide $Q$-factors between 1 and $2 \times 10^6$ at room temperature. 

\begin{figure}[htb]
    \label{fig:Rakon}
\begin{minipage}{0.33\linewidth}
    \includegraphics[width=0.90\linewidth]{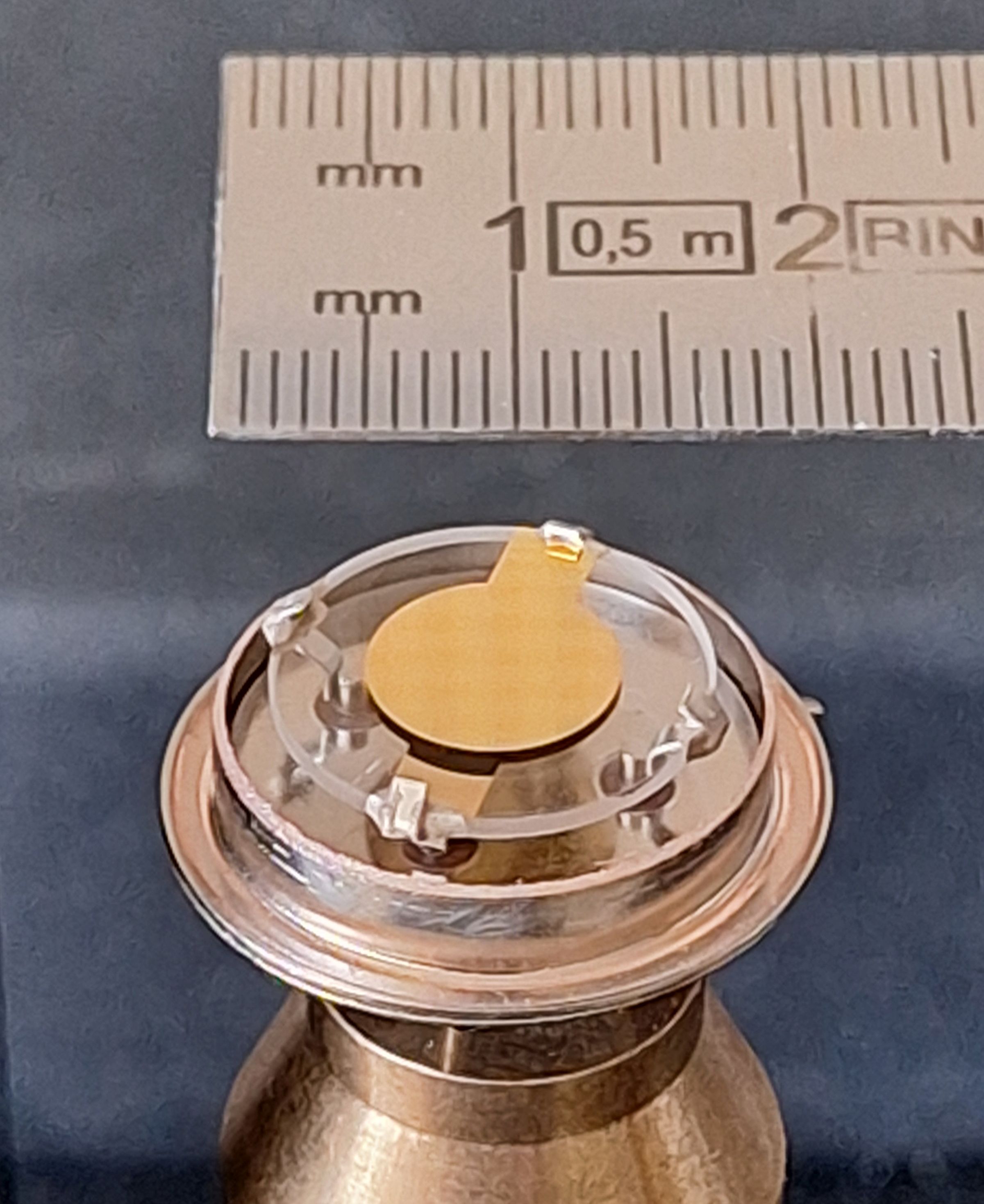}
\end{minipage}
\begin{minipage}{0.66\linewidth}
    \vspace{3.8cm}
    \caption{Picture of a 1~mm thick, 14 mm diameter BAW sample made by a plano-convex quartz crystal anchored by four clamps and thin gold electrodes deposited on the surfaces (printed with permission from Rakon Ltd).}
    \label{fig:Rakon}
\end{minipage}
\end{figure}

\acapo The curvature radius is smaller in the reference ($R_c = 300$~mm) than in the new samples ($R_c = 600$~mm). Consequently, the trapping of the acoustic modes is less pronounced in the new samples, and the vibrations extend closer to the edges of the crystals. 
However, as illustrated in Fig.~\ref{fig:Modes}, the main modes are well trapped in both geometries for $n\geq3$. The mode amplitudes have 2D-normal profiles of comparable radial width, $\sigma_n = {(d^3R_c)}^{1/4} / \sqrt{\pi n}$ ($\sim 2.5$~mm for $n=1$), and effective mass, given by the integral over the crystal volume of the amplitude profile times density.



\acapo These profiles are derived under the approximation of an isotropic crystal lattice and a cavity with thickness $d=d(x,y)$, slowly varying as a function of the radial coordinate~\cite{Tiersten86, Goryachev_2014}. The surface curvature effectively acts as a two-dimensional harmonic potential because of the inverse proportionality between the frequency (phonon energy) and the local plate thickness. The maximum thickness determines the frequency
of the main modes, as specified by Eq.~(\ref{eq:modes}). In addition, the plano-convex geometry admits thickness modes with polynomial-normal amplitude profiles. In this case, the amplitude maximum is off the disk centre. Consequently, these modes are frequency-shifted relative to the main modes:
\begin{equation}
    \label{eq:freq-shift}
    f_\lambda \sim n f_{\{X,1,0,0\}} + \frac{n-1}{2}(\Delta f_x+\Delta f_y) + m \Delta f_x + p \Delta f_y,
\end{equation}
where $\lambda=\{X,n,m,p\}$ identifies the mode, through its polarization, overtone number, and two additional integer numbers, $m$ and $p$, that characterize the mode amplitude in $(x,y)$. The frequency shifts $\Delta f_x$ and $\Delta f_y$ depend on the crystal properties and curvature, and their measurement quantifies the trapping in the $x$ and $y$ directions~\cite{Goryachev_2014}. The parameterization includes two directions because the crystal lattice is not symmetric around the vertical axis. These modes are more spread in the transverse plane, less trapped than the main modes ($m=p=0$), and generally result in lower $Q$-factors. 

\begin{figure}[htb!]
    \centering
    \raisebox{0.cm}{\includegraphics[width=0.47\linewidth]{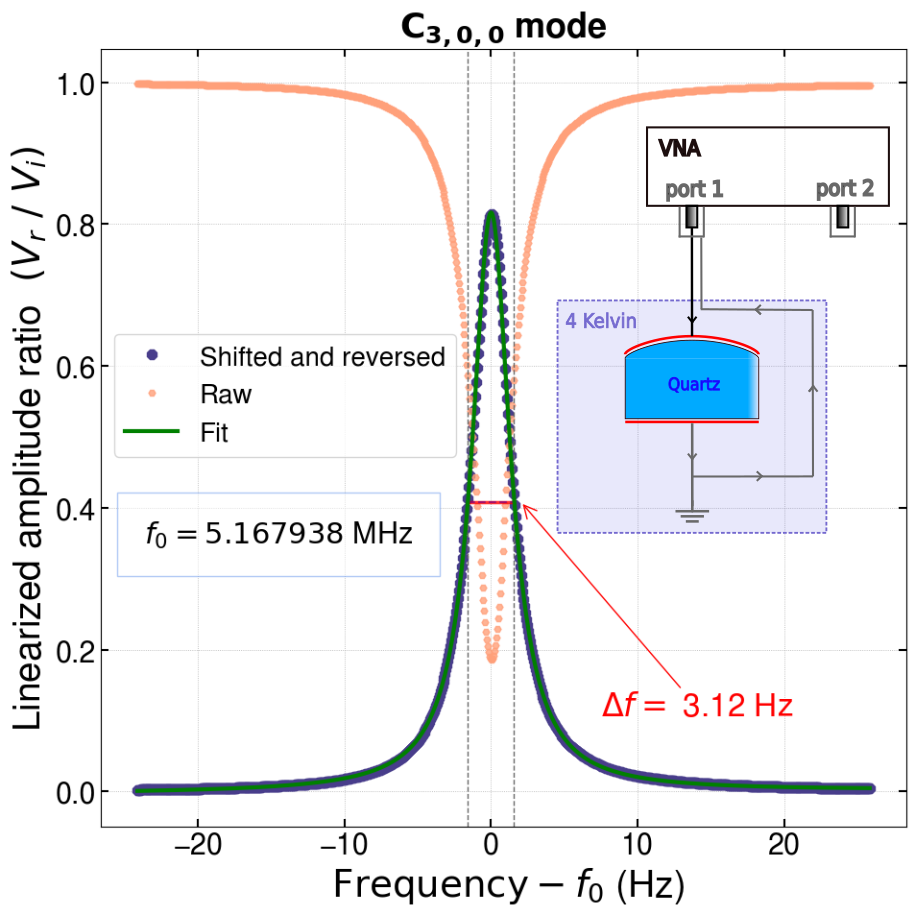}}
    \includegraphics[width=0.52\linewidth]{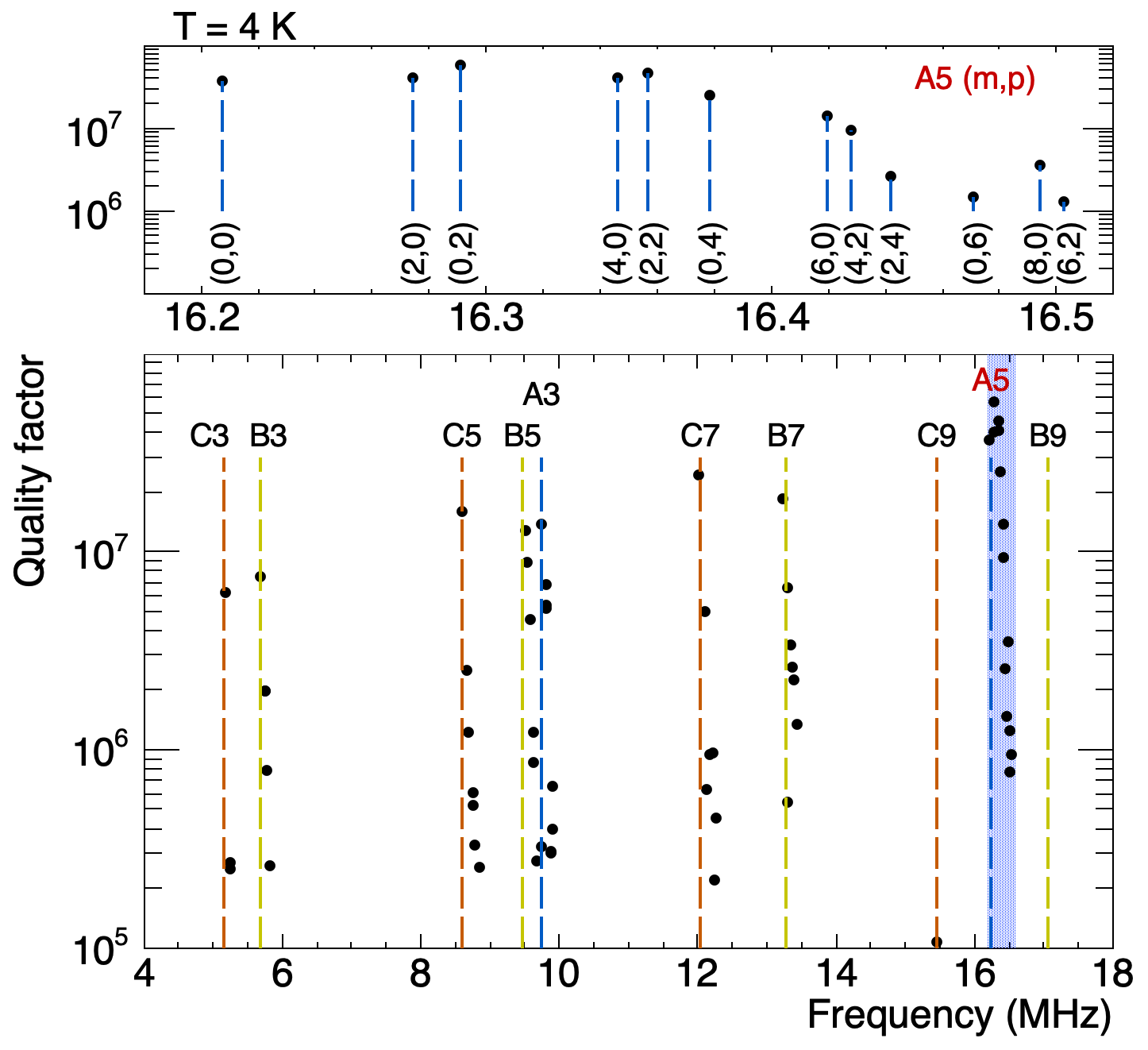}
    \caption{{\it Left}: Frequency scan around the resonance for the mode $\lambda = \{C,3,0,0\}$ at $T=3.5$~K. The inset shows the connection to the VNA. {\it Right}: Identified resonant modes for one BAW sample at $T=4$~K as a function of the frequency: the main modes ($m=p=0$) are marked with vertical lines; the top panel shows the frequency pattern of thickness transverse modes for $\lambda = \{A,5,m,p\}$.}
    \label{fig:modeScan}
\end{figure}

\acapo The resonator characterization relies on impedance measurements performed through a calibrated Vector Network Analyzer (VNA). For example, the VNA can be coupled to the resonator in a reflection geometry, with one electrode connected to one port and the other to ground. The coupling coefficient $\beta_{\lambda} = R_{\lambda} / R_{\mathrm{VNA}}$ can be extracted from the values of the scattering parameters on resonance, where $R_{\lambda}$ is the motional resistance of mode and $R_{\mathrm{VNA}} = \SI{50}{\ohm}$ is the output impedance of the VNA. A typical resonance scan is shown in the left panel of Fig.~\ref{fig:modeScan}. The quality factor extracted from the full width at half maximum of the resonance curve, $Q_{\lambda,L} = f_{\lambda} / \Delta f$ includes both resonator losses and those due to coupling with the VNA transmission line (`loading' effect). The quality factors of the resonator ('unloaded') are derived from the unfolding of the coupling coefficient $Q_{\lambda, u} = (1 + \beta_{\lambda}) Q_{\lambda, L}$. 

\acapo The unloaded $Q_\lambda$ factors of the resonances observed between 5 and 20~MHz at $T=3.5$~K with one of the new samples are displayed in the right panel of Fig.~\ref{fig:modeScan}. Families with overtone numbers $n=3$,~5, and 7 are well visible for the longitudinal, fast, and slow shear modes. The resonant frequencies are approximately 0.15\% lower than at room temperature, indicating a slight variation in the phase velocity with temperature. For all the main modes ($m=p=0$), the $Q$-factors at low temperature are over $10^7$, and reach $3.5\times10^7$ for the fifth overtone of the longitudinal mode, $\lambda=\{A, 5, 0, 0\}$. Within a family, the $Q$-factors decrease with increasing frequency, indicating that for $(m,p)\neq(0,0)$ the modes are less trapped as predicted. The inset in the figure shows a magnification of the frequency pattern for $\lambda = \{A,5,m,p\}$. Only modes with even $m$ and $p$ exhibit a piezoelectric counterpart. 

\begin{figure}[hbt]
    \centering
    \raisebox{-0.1mm}{\includegraphics[width=0.491\linewidth]{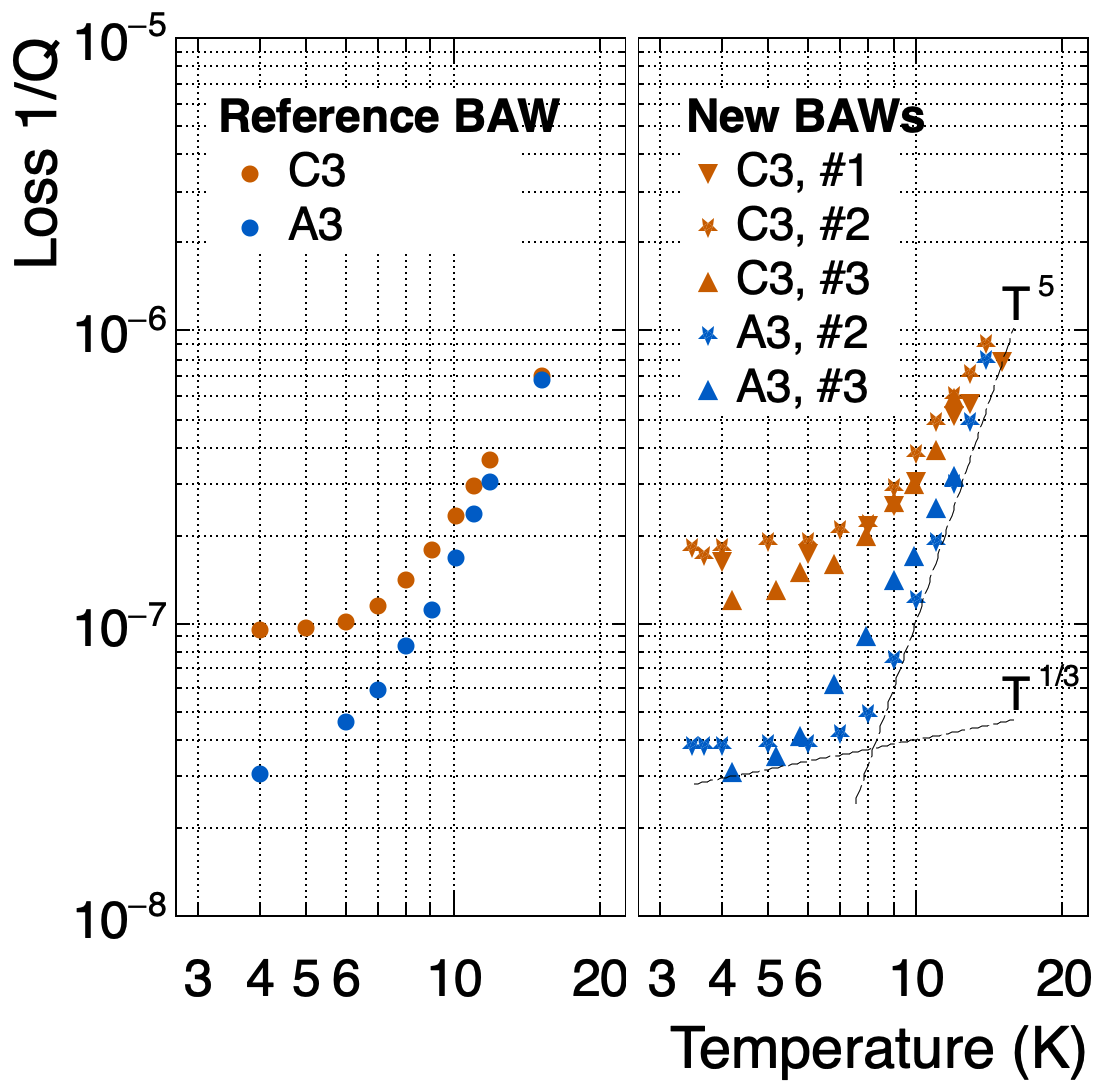}}
    \includegraphics[width=0.495\linewidth]{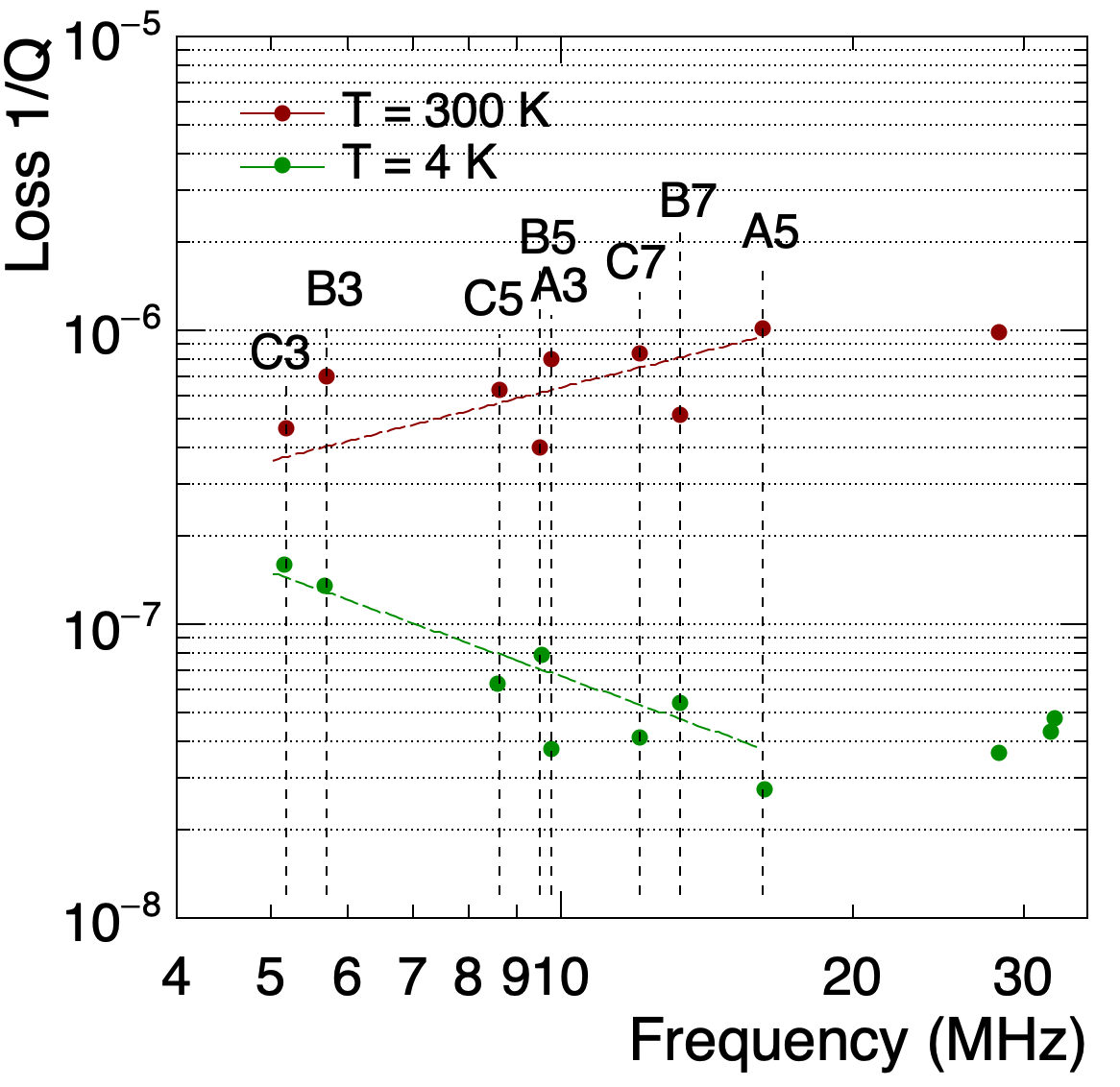}
    \caption{{\it Left}: Dependence of the resonator losses on temperature for the modes with $\lambda = \{X,3,0,0\}$; the dashed lines represent the two-level-system loss regime ($T^{1/3}$), and higher-order scaling regimes ($T^{2.8}$ and $T^{5.7}$), ascribed to phonon-phonon scattering. {\it Right}: Measurements of $1/Q$ as a function of the frequency for the observed main modes $\lambda=\{X,n,0,0\}$ ($n=3, 5, 7$) at $T = 4$ and 300~K; the dashed lines are empirical power laws through the points with indices ${0.8}$ and ${-1.2}$, respectively.}
    \label{fig:LossesVsTemp}
\end{figure}

\acapo The gain in performance at low temperature is remarkable and comparable to the reference and MAGE resonators~\cite{SciRep13-10638}. 
To gain insight into loss mechanisms in our samples, we studied the temperature dependence of the cavity losses. Temperature-dependent intrinsic losses, $1/Q_{T}$, and mechanical losses such as the lack of trapping or losses through the electrodes, $1/Q_m$, contribute to the total loss, $1/Q = 1/Q_{T}+1/Q_{m}$.
Results are shown in the left panel of Fig.~\ref{fig:LossesVsTemp} for a few representative modes with $m=p=0$ for both the reference and the new samples. The longitudinal modes benefit most from low-temperature operation. The losses follow a power law with index $\gamma \gtrsim 3$ down to $T\sim 7$~K. At a lower temperature, the losses show a milder scaling as $\propto T^{1/3}$. These behaviours agree with previous studies on BAWs, including the reference sample of this study~\cite{Galliou2013, campbell2025}. The main loss mechanism is attributed to the scattering of acoustic waves off thermal phonons. The gentler low-temperature dependence is linked to phonon interactions with two-level systems related to ionic impurities within the crystal. 
From $T \sim 3$~K to $\sim 20$~mK, the $Q$-factors of the new samples improve only marginally, suggesting that in this regime the performance is limited by mechanical losses. 



\acapo Thermal losses in acoustic resonators have a complex dependency on frequency and temperature, contingent on the relationship between thermal and acoustic phonon energies. For the main modes at low overtones, the loss versus frequency follows approximately a power law, with power indices $0.8$ and $-1.2$ at room temperature and $T=4$~K, respectively (right panel of Fig.~\ref{fig:LossesVsTemp}). Not all modes are observed above 16 MHz ($A5$), and trends become less clear. Residual mechanical losses through the clamps and electrodes may explain the observed pattern~\cite{Galliou2013, campbell2025}. Yet, resonators with the electrodes deposited on the crystal have a simple design and are easier to realise than electrode-separated devices. They represent a good choice for future developments, as they already provide excellent performance on the modes with significant coupling to gravitational waves. 

\section{Predicted sensitivities and optimizations}
\label{Sec3}


The new BAW samples are a good option for building a prototype antenna for high-frequency gravitational waves: the effective mass is similar to the reference sample, despite the smaller transverse size, and the surface curvature provides effective trapping of the main acoustic modes, with observed $Q$-factors above $10^7$. In the range between 5 and 20~MHz, these resonators provide a strain sensitivity comparable to the MAGE experiment.

\begin{figure}[htb]
    \centering
    \includegraphics[width=0.80\linewidth]{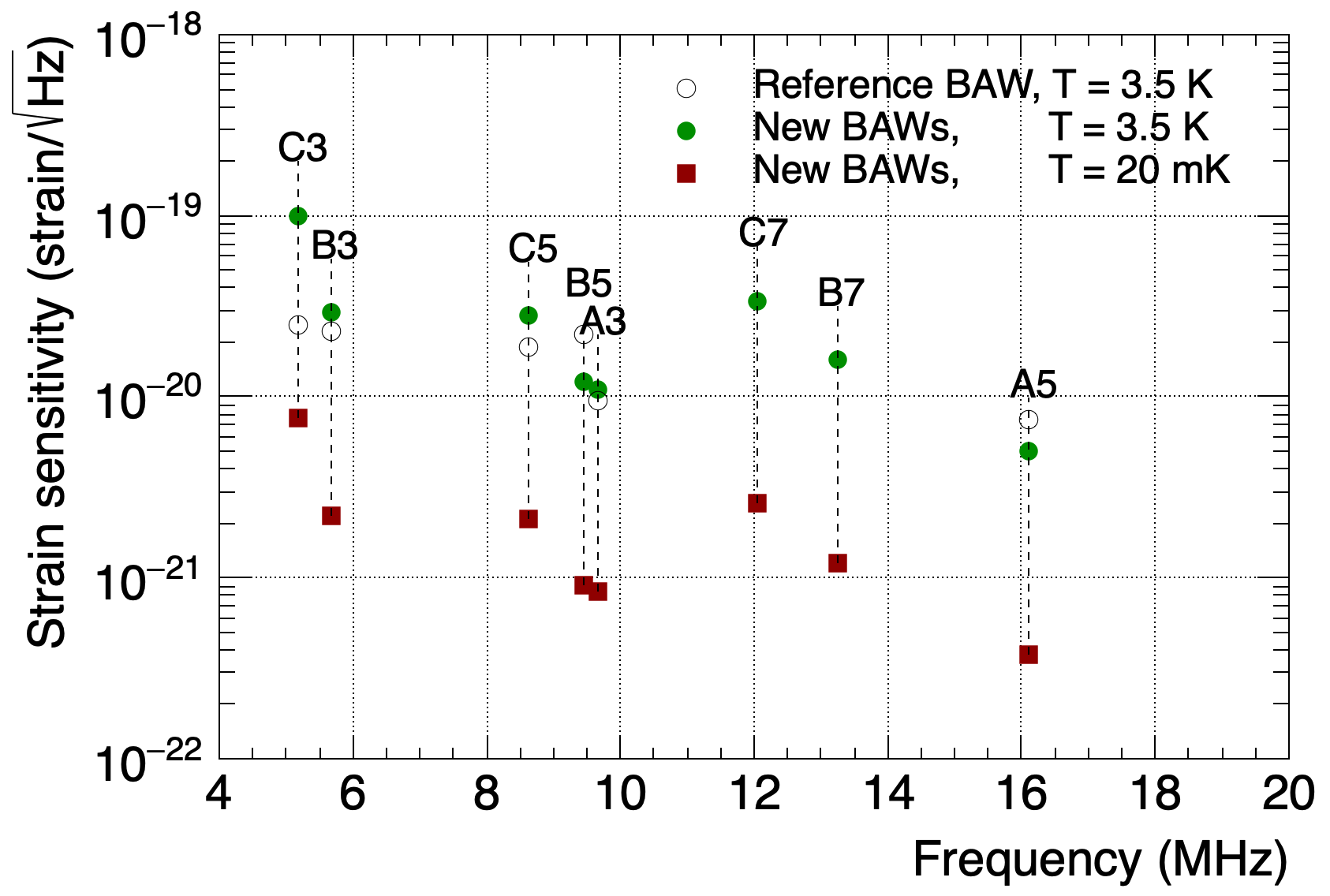}
    \caption{Predicted single-sensor peak spectral strain sensitivity for multiple modes at $T=3.5$~K for the reference (empty dots) and the new samples (full dots), and for the new samples at $T = 20$~mK (squares).}
    \label{fig:PeakStrain}
\end{figure}

\acapo The results of our characterization are summarized in Fig.~\ref{fig:PeakStrain}, showing the single-sided peak spectral strain sensitivity for multiple BAW modes and different devices at $T=3.5$~K. The new BAW samples (full dots) and the MAGE reference sample (open dots) provide comparable single-sensor peak strain sensitivities around $1$--$3\times 10^{-20}\,\textrm{strain}/\sqrt{\textrm{Hz}}$. The predicted sensitivities are scaled from the sensitivity reported by the MAGE experiment~\cite{SciRep13-10638}, based on the $Q_\lambda$ measurements, the effective mass, and phonon trapping described in this paper. The mode temperatures, $T_\lambda$, are assumed equal to the operating temperature, because the characterisation of the readout chain and calibration of the SQUIDs are still in progress. 

\acapo The figure also shows the single-sensor sensitivity predicted for operation of the same BAW samples at $T=20$~mK (squares). The thermal noise reduction results in a tenfold increase in sensitivity, achieving values around or below $10^{-21}\,\textrm{strain}/\sqrt{\textrm{Hz}}$, despite $Q$-factors not showing significant improvement below 3.5~K. 

Commercially available resonators only provide specific frequencies related to telecommunication or clock standards. Additionally, efforts toward miniaturization shift the reference standards to higher frequencies, resulting in reduced coupling to the incoming gravitational waves due to the lower mass of the resonator. Consequently, custom BAWs are being developed to enhance the sensitivity and extend the frequency coverage at low frequencies. 
Assuming complete trapping of the phonons by proper disk shaping, considering the proportionality of the effective mass with the crystal density $\rho$ and volume $\pi L^2 d$, and expressing the frequency according to Eq.~(\ref{eq:modes}), the peak strain sensitivity scales as: 
\begin{equation}
    \label{eq:scalingSens}
        S^{+}_{h}(\omega_{\lambda}) \propto
        \frac{n}{L}
\sqrt{\frac{k_bT_{\lambda}}{\rho Q_{\lambda}v^3}} \ \ \ [\textrm{strain}/\sqrt{\textrm{Hz}}].
\end{equation}

\acapo Optimized performance requires a large radius, which impacts the resonator mass, a fast crystal, and low overtone numbers. Instead, the crystal thickness determines the BAW frequency spectrum but does not directly influence sensitivity, which depends on frequency solely through the overtone number. Therefore, a broad frequency coverage while maintaining consistent sensitivity requires several BAW devices with varying thicknesses. 


\acapo
For the project's second phase, we are developing high-purity quartz BAWs of eight different thicknesses varying from 2 to 27~mm, as illustrated in Fig.~\ref{fig:setup}. We are working closely with manufacturers to address several aspects of crystal customization, including cutting, grinding, polishing, and annealing. Although resonators of 1~kg and 150~mm diameter have been deemed feasible for metrology applications~\cite{6416499}, we limit the crystal size to 25~mm diameter for initial production, as the strict tolerance requirements make the manufacturing of large SC-cut crystals challenging. In contrast, AT-cut resonators have less stringent requirements but still necessitate low-temperature characterization. Electrodes deposited on the disk surfaces provide a simple and effective solution for signal pick-up, although alternative packages are also under investigation. The results of these developments will be discussed in future publications. 
The primary objective is to achieve high quality factors in large crystals at low overtones, chiefly for the mode $\lambda=\{A,1,0,0\}$, which holds the promise of the highest sensitivity to gravitational waves for quartz BAWs.
The envelope of the projected peak strain sensitivities of an array of two sets of eight BAWs of these dimensions is shown in Fig.~\ref{fig:sensitivity}, where a fivefold gain on the current sensitivity is anticipated from performance optimizations at low overtones and the increase in the detector mass, including resonator diameter and total quantity.

\section*{Acknowledgements}
This research is developed within the framework of the ``{\it Centro Bicocca di Cosmologia Quantitativa}'' (BiCoQ), supported by the Italian Ministry for Universities and Research (MUR) under Grant ``{\it Dipartimenti di Eccellenza 2023-2027}''. The PNRR MUR Project also provided support to this work under grant PE0000023-NQSTI. We are indebted and sincerely grateful to Mike Tobar, Maxim Goryachev, and William Campbell for their assistance and guidance in launching this activity. We thank Mauro Fasoli and Chiara Liliana Boldrini, from the Department of Material Science at the University of Milano Bicocca, for the geometry, roughness, and crystal properties measurements of the 'sacrificed' BAW sample. 


\raggedright
\bibliographystyle{ieeetr}
\bibliography{sample} 

\begin{thebibliography}{10}

\bibitem{PhysRevLett.116.061102}
B.~P. Abbott {\em et~al.}, ``Observation of gravitational waves from a binary black hole merger,'' {\em Phys. Rev. Lett.}, vol.~116, p.~061102, Feb 2016.

\bibitem{Abbott2017}
B.~Abbott {\em et~al.}, ``A gravitational-wave standard siren measurement of the {H}ubble constant,'' {\em Nature}, vol.~551, pp.~85--88, Nov 2017.

\bibitem{Moore_2015}
C.~J. Moore, R.~H. Cole, and C.~P.~L. Berry, ``Gravitational-wave sensitivity curves,'' {\em Classical and Quantum Gravity}, vol.~32, p.~015014, Dec. 2014.

\bibitem{Aggarwal_2021}
N.~Aggarwal {\em et~al.}, ``Challenges and opportunities of gravitational-wave searches at {MH}z to {GH}z frequencies,'' {\em Living Reviews in Relativity}, vol.~24, Dec 2021.

\bibitem{PhysRevD.106.103520}
G.~Franciolini, A.~Maharana, and F.~Muia, ``Hunt for light primordial black hole dark matter with ultrahigh-frequency gravitational waves,'' {\em Phys. Rev. D}, vol.~106, p.~103520, Nov 2022.

\bibitem{PhysRevD.83.044026}
A.~Arvanitaki and S.~Dubovsky, ``Exploring the string axiverse with precision black hole physics,'' {\em Phys. Rev. D}, vol.~83, p.~044026, Feb 2011.

\bibitem{casalderreysolana2022}
J.~Casalderrey-Solana, D.~Mateos, and M.~Sanchez-Garitaonandia, ``Mega-{H}ertz gravitational waves from neutron star mergers,'' 2022.

\bibitem{Cruise_2006}
A.~M. Cruise and R.~M.~J. Ingley, ``A prototype gravitational wave detector for 100 {MH}z,'' {\em Classical and Quantum Gravity}, vol.~23, p.~6185, oct 2006.

\bibitem{Cruise_2012}
A.~M. Cruise, ``The potential for very high-frequency gravitational wave detection,'' {\em Classical and Quantum Gravity}, vol.~29, p.~095003, apr 2012.

\bibitem{PhysRevD.77.022002}
A.~Nishizawa, S.~Kawamura, T.~Akutsu, K.~Arai, K.~Yamamoto, D.~Tatsumi, E.~Nishida, M.-a. Sakagami, T.~Chiba, R.~Takahashi, and N.~Sugiyama, ``Laser-interferometric detectors for gravitational wave backgrounds at 100 {MH}z: Detector design and sensitivity,'' {\em Phys. Rev. D}, vol.~77, p.~022002, Jan 2008.

\bibitem{PhysRevLett.101.101101}
T.~Akutsu, S.~Kawamura, A.~Nishizawa, K.~Arai, K.~Yamamoto, D.~Tatsumi, S.~Nagano, E.~Nishida, T.~Chiba, R.~Takahashi, N.~Sugiyama, M.~Fukushima, T.~Yamazaki, and M.-K. Fujimoto, ``Search for a stochastic background of 100-{MH}z gravitational waves with laser interferometers,'' {\em Phys. Rev. Lett.}, vol.~101, p.~101101, Sep 2008.

\bibitem{PhysRevD.95.063002}
A.~S. Chou, R.~Gustafson, C.~Hogan, B.~Kamai, O.~Kwon, R.~Lanza, S.~L. Larson, L.~McCuller, S.~S. Meyer, J.~Richardson, C.~Stoughton, R.~Tomlin, and R.~Weiss, ``{MH}z gravitational wave constraints with decameter {M}ichelson interferometers,'' {\em Phys. Rev. D}, vol.~95, p.~063002, Mar 2017.

\bibitem{Ejlli2019}
A.~Ejlli, Ejlli, A.~D., Cruise, {\em et~al.}, ``Upper limits on the amplitude of ultra-high-frequency gravitational waves from graviton to photon conversion,'' {\em The European Physical Journal C}, vol.~79, p.~1032, 2019.

\bibitem{PhysRevD.102.103501}
T.~Fujita, K.~Kamada, and Y.~Nakai, ``Gravitational waves from primordial magnetic fields via photon-graviton conversion,'' {\em Phys. Rev. D}, vol.~102, p.~103501, Nov 2020.

\bibitem{Ito_2020}
A.~Ito, T.~Ikeda, K.~Miuchi, and J.~Soda, ``Probing {GH}z gravitational waves with graviton–magnon resonance,'' {\em The European Physical Journal C}, vol.~80, Feb 2020.

\bibitem{Herman_2021}
N.~Herman, A.~Fűzfa, L.~Lehoucq, and S.~Clesse, ``Detecting planetary-mass primordial black holes with resonant electromagnetic gravitational-wave detectors,'' {\em Physical Review D}, vol.~104, July 2021.

\bibitem{Tobar_2022}
M.~E. Tobar, C.~A. Thomson, W.~M. Campbell, A.~Quiskamp, J.~F. Bourhill, B.~T. McAllister, E.~N. Ivanov, and M.~Goryachev, ``Comparing instrument spectral sensitivity of dissimilar electromagnetic haloscopes to axion dark matter and high frequency gravitational waves,'' {\em Symmetry}, vol.~14, p.~2165, Oct 2022.

\bibitem{Domcke_2022}
V.~Domcke, C.~Garcia-Cely, and N.~L. Rodd, ``Novel search for high-frequency gravitational waves with low-mass axion haloscopes,'' {\em Physical Review Letters}, vol.~129, Jul 2022.

\bibitem{PhysRevD.110.023018}
C.~Gatti, L.~Visinelli, and M.~Zantedeschi, ``Cavity detection of gravitational waves: Where do we stand?,'' {\em Phys. Rev. D}, vol.~110, p.~023018, Jul 2024.

\bibitem{GAO20242795}
Y.~Gao, W.~Xu, and H.~Zhang, ``A {M}össbauer scheme to probe gravitational waves,'' {\em Science Bulletin}, vol.~69, no.~18, pp.~2795--2798, 2024.

\bibitem{domcke2024magnetsweberbargravitational}
V.~Domcke, S.~A.~R. Ellis, and N.~L. Rodd, ``Magnets are {W}eber bar gravitational wave detectors,'' 2024.

\bibitem{PhysRevLett.110.071105}
A.~Arvanitaki and A.~A. Geraci, ``Detecting high-frequency gravitational waves with optically levitated sensors,'' {\em Phys. Rev. Lett.}, vol.~110, p.~071105, Feb 2013.

\bibitem{PhysRevD.90.102005}
M.~Goryachev and M.~E. Tobar, ``Gravitational wave detection with high-frequency phonon trapping acoustic cavities,'' {\em Phys. Rev. D}, vol.~90, p.~102005, Nov 2014.
\newblock Erratum: {\it Phys. Rev. D}, 108.129901, Dec, 2023.

\bibitem{tobar2024G}
G.~Tobar, I.~Pikovski, and M.~E. Tobar, ``Detecting k{H}z gravitons from a neutron star merger with a multi-mode resonant mass detector,'' 2024.

\bibitem{PhysRevLett.127.071102}
M.~Goryachev, W.~M. Campbell, I.~S. Heng, S.~Galliou, E.~N. Ivanov, and M.~E. Tobar, ``Rare events detected with a bulk acoustic wave high frequency gravitational wave antenna,'' {\em Phys. Rev. Lett.}, vol.~127, p.~071102, Aug 2021.

\bibitem{SciRep13-10638}
W.~Campbell, M.~Goryachev, and M.~E. Tobar, ``The {M}ulti-mode {A}coustic {G}ravitational wave {E}xperiment: {MAGE},'' {\em Scientific Reports}, vol.~13, p.~10638, Jun 2023.

\bibitem{PhysRevLett.116.031102}
A.~Arvanitaki, S.~Dimopoulos, and K.~Van~Tilburg, ``Sound of dark matter: Searching for light scalars with resonant-mass detectors,'' {\em Phys. Rev. Lett.}, vol.~116, p.~031102, Jan 2016.

\bibitem{PhysRevLett.124.151301}
J.~Manley, D.~J. Wilson, R.~Stump, D.~Grin, and S.~Singh, ``Searching for {S}calar {D}ark {M}atter with {C}ompact {M}echanical {R}esonators,'' {\em Phys. Rev. Lett.}, vol.~124, p.~151301, Apr 2020.

\bibitem{trickle2025piezoelectricbulkacousticresonators}
T.~Trickle, ``Piezoelectric bulk acoustic resonators for dark photon detection,'' 2025.

\bibitem{Goryachev_2014}
M.~Goryachev, E.~N. Ivanov, F.~van Kann, S.~Galliou, and M.~E. Tobar, ``Observation of the fundamental {N}yquist noise limit in an ultra-high {Q}-factor cryogenic bulk acoustic wave cavity,'' {\em Applied Physics Letters}, vol.~105, Oct 2014.

\bibitem{Tiersten86}
D.~Stevens and H.~Tiesten, ``An analysis of doubly rotated quartz resonators utilizing essentially thickness modes with transverse variation,'' {\em The Journal of the Acoustical Society of America}, vol.~79, pp.~1811--1826, 1986.

\bibitem{Galliou2013}
S.~Galliou, M.~Goryachev, R.~Bourquin, P.~Abbé, J.~P. Aubry, and M.~E. Tobar, ``Extremely low loss phonon-trapping cryogenic acoustic cavities for future physical experiments,'' {\em Scientific Reports}, vol.~3, p.~2132, Jul 2012.

\bibitem{1537081}
R.~Besson, ``A new "electrodeless" resonator design,'' in {\em 31st Annual Symposium on Frequency Control}, pp.~147--152, 1977.

\bibitem{campbell2025}
W.~M. Campbell, L.~Mariani, S.~Parashar, M.~E. Tobar, and M.~Goryachev, ``Low temperature properties of low-loss macroscopic lithium niobate bulk acoustic wave resonators,'' 2025.

\bibitem{6416499}
J.~Vig and D.~Howe, ``A one-kilogram quartz resonator as a mass standard,'' {\em IEEE Transactions on Ultrasonics, Ferroelectrics, and Frequency Control}, vol.~60, no.~2, pp.~428--431, 2013.

\end{thebibliography}

\end{document}